\documentstyle[12pt,psfig]{article}
\def\spose#1{\hbox to 0pt{#1\hss}} %
\def\gta{\mathrel{\spose{\lower 3pt\hbox{$\mathchar"218$}}
      \raise 2.0pt\hbox{$\mathchar"13E$}}}
\def\lta{\mathrel{\spose{\lower 3pt\hbox{$\mathchar"218$}}
      \raise 2.0pt\hbox{$\mathchar"13C$}}}
\begin{document}

\begin{center}
{\bf Material enhancement in protoplanetary nebulae \\
by particle drift through evaporation fronts \\}
\vspace{0.25 in}
Jeffrey N. Cuzzi and Kevin J. Zahnle \\
Space Science Division, Ames Research Center\\
Submitted to Astrophysical Journal, March 29, 2004\\
Revised June 10, 2004 and September 8, 2004\\
In press, Vol 614, October 10 2004 issue\\
\vspace{3.0 in}
\end{center}

\begin{obeylines}
Mail Stop 245-3, Ames Research Center, NASA; Moffett Field, CA 94035
jcuzzi@mail.arc.nasa.gov; kzahnle@mail.arc.nasa.gov
\vspace{0.25 in}
Short title: Evaporation fronts in protoplanetary nebulae 
\vspace{0.25 in}
Keywords: solar system: formation; accretion, accretion disk; 
(stars):planetary systems:protoplanetary disk; turbulence; diffusion
\vspace{0.25 in}
26 double-spaced pages including figures and references; 3 figures, 1 Table
\end{obeylines}

\newpage
{ \bf
\begin{center}
Abstract
\end{center}
Solid material in a protoplanetary nebula is subject to vigorous redistribution
processes relative to the nebula gas. Meter-sized particles drift rapidly
inwards near the nebula midplane, and material evaporates when the particles
cross a condensation/evaporation boundary. The material cannot be removed as
fast in its vapor form as it is being supplied in solid form, so its
concentration increases locally by a large factor (more than an order of
magnitude under nominal conditions). As time goes on, the vapor phase
enhancement propagates for long distances inside the evaporation boundary
(potentially all the way in to the star). Meanwhile, material is enhanced in
its solid form over a characteristic lengthscale outside the evaporation
boundary. This effect is applicable to any condensible (water, silicates, {\it
etc.}). Three distinct radial enhancement/depletion regimes can be discerned by
use of a simple model. Meteoritics applications include oxygen fugacity and
isotopic variations, as well as isotopic homogenization in silicates. Planetary
system applications include more robust enhancement of solids in Jupiter's core
formation region than previously suggested. Astrophysical applications include
differential, time-dependent enhancement of vapor phase CO and H$_2$O in the
terrestrial planet regions of actively accreting protoplanetary disks.}

\vspace{1.0 in}
\newpage

\section{Introduction}

Matter doesn't simply condense from a cooling protoplanetary nebula at its
cosmic relative abundance and remain in place. Significant inward radial
transport of solids occurs relative to nebula hydrogen (Morfill and
V{\"o}lk 1984, Stepinski and Valageas 1997), and trace vapors migrate outwards
to condensation fronts (Stevenson and Lunine 1988). Each compound has its own
condensation front; water's is often called the snowline. Previous work has
stressed the role that condensation of outwardly diffused vapor plays in
enhancing the density of solids at the snowline (Stevenson and Lunine 1988).
However, inward particle drift can also enhance the abundance of a vapor inside
the condensation/evaporation boundary (Cuzzi et al 2003). Hence, given
turbulent mixing, the density of solids just outside the boundary also grows.
That particle drift can have this effect has been recognized (Morfill and
V{\"o}lk 1984), but the magnitude of the effect has not been recognized. Here
we show that, in general, inward particle drift is more effective than outward
vapor diffusion at enhancing solids outside the condensation/evaporation
boundary, and that in many cases of interest vapor is also strongly enhanced
{\it interior to} the condensation/evaporation boundary. To emphasize the
difference between this new process and previous work, we will often refer to
the condensation/evaporation boundary as the {\it evaporation front}, which more
accurately captures the directionality of the process described here.

In this paper we present a minimal model that illustrates the key physical
arguments. We divide a condensible solid into three size classes distinguished
by their transport properties: vapor and small grains that are tightly coupled
to the movements of the gas; large bodies that orbit unaffected by gas drag; and
mid-sized particles (boulders, or rubble --- typically on the order of a meter)
that are strongly affected by both gravity and gas drag (Weidenschilling 1977).
The latter can drift orders of magnitude faster than the nebula gas and carry a
net flux greatly exceeding that which is coupled to the gas. The particles
evaporate when they reach the evaporation front, enhancing the abundance of the
condensible in the vapor phase (Cuzzi et al 2003). While the physics is
applicable to volatiles in general, we focus here on water as a volatile of
special interest. The enhanced water vapor abundance spreads radially inwards
from the evaporation front on a timescale which is short compared to the
lifetime of the nebula, potentially determining the mineralogy of primitive
meteorites. Planetesimal growth just outside the condensation/evaporation
boundary, or snowline, provides a sink that ultimately depletes the vapor at
all locations inside the boundary (Stevenson and Lunine 1988). Ice enhancement
outside the snowline can influence the location and timescales of giant planet
core formation (Morfill and V{\"o}lk 1984, Stevenson and Lunine 1988). Our
model illustrates all these regimes of behavior.

\subsection{Nebula evolution and turbulence:} Protoplanetary disks evolve; both
their surface mass density $\sigma_g$ and mass accretion rate $\dot{M} = 2 \pi R
\sigma_g V_n$ (where $V_n$ is the nebula gas advection velocity) decrease with
time over a period of several Myr (Calvet et al 2000). The disk is heated by
gravitational energy release ($G\dot{M}/R$) and illumination from the star. In
recent models of actively accreting disks (Bell et al, 1997, Stepinski 1998;
{\it cf.} also Woolum and Cassen 1999), midplane temperatures are hot enough in
the terrestrial region to vaporize common silicates at early times, and to
vaporize water in the jovian region over a more extended period. The physical
cause of nebula evolution remains problematic; turbulent viscosity is now
thought to face difficulties (Stone et al 2000). However, turbulence can exist,
providing diffusivity, without necessarily providing the viscosity needed to
evolve the disk (Prinn 1990; Cuzzi et al 2001). Since diffusivity, rather than
viscosity, is of prime interest here, we will assume nebulae which are weakly
turbulent. As discussed below and by Cuzzi and Weidenschilling (2004),
turbulence plays several roles: it diffuses grains and vapor down concentration
gradients, often against the flow of nebular drift (Cuzzi et al 2003); it
frustrates the growth of particles beyond about a meter in size; and it
determines the midplane particle density (and thus particle and planetesimal
growth rates).

We presume a weakly turbulent nebula with effective viscosity $\nu_T = \alpha c
H$, where $c$ is the sound speed, $H$ is the scale height, and the parameter
$\alpha$ is defined by the evolutionary mass accretion rate of the nebula:
$\alpha \equiv \dot{M}/3 \pi \sigma_g c H$. Hence the advection velocity $V_n =
3\alpha c H/2R$. Observations and models typically suggest that $\alpha \approx
10^{-5} - 10^{-2}$ and $H\approx R/20$. The turbulent diffusivity ${\cal D} =
\nu_T/Pr_t$ is related to the viscosity by the turbulent Prandtl number, $Pr_t$. It is
usually presumed that $Pr_t=1$ in a turbulent nebula, but in general $Pr_t\neq 1$;
e.g.\ Prinn (1990) suggests that $Pr_t \ll 1$ and that mixing is efficient even
if the nebula evolves slowly. The characteristic velocity of large eddies is
$V_g \sim c (\alpha/Pr_t)^{1/2}$. Nebular evolution is driven by $\nu_T$, while
${\cal D}$ describes mixing.

\subsection{Turbulence and particle growth to meter-size:} Particle growth is
easy up to meter-size, but problematic beyond. With or without turbulence, the
relative velocities between sub-m-size particles are low and the first
aggregates probably grow by simple sticking into porous, dissipative structures
(Weidenschilling and Cuzzi 1993; Cuzzi and Weidenschilling 2004). Most growth
occurs as large particles sweep up smaller ones (Weidenschilling 1997). Under
nonturbulent conditions, large particles sink into a high density midplane
layer in which relative velocities are low, and subsequent growth to
planetesimal sizes is rapid (Weidenschilling 1997, Cuzzi et al 1993). In
turbulence, because meter-sized particles couple to the largest eddies and
achieve random velocities on the order of $V_g$, they remain in a layer of
finite thickness $h_L \sim V_g/\Omega_K \sim H (\alpha/Pr_t)^{1/2}$. Even weak
turbulence ($\alpha > 10^{-6}$, $Pr_t=1$) keeps the density of the midplane layer
and the ensuing particle growth rate substantially below their nonturbulent
values (Dubrulle et al 1995). Furthermore, meter-sized particles collide with
each other at speeds comparable to $V_g$ - meters to tens of m/sec - probably
fragmenting into their smaller constituents. While particles approaching
meter-size, having impact strength of $10^6$ erg/cm$^3$ (Sirono and Greenberg
2000), can survive mutual collisions in turbulence of $\alpha = 10^{-4}$ (Cuzzi
and Weidenschilling 2004), one suspects that further {\it incremental growth}
may stall at the m-size limit, at least while turbulence this large persists.
Other physics, however, may come into play ({\it eg.}, Cuzzi et al 2001).
Detailed models of incremental growth, with realistic sticking and erosion
based on laboratory experiments, tend to quickly produce broad power law size
distributions which contain equal mass per decade of particle radius
(Weidenschilling 1997, 2000). Assuming that growth beyond the meter-size range
is frustrated as described above, and conservatively assuming ten decades of
particle size (microns to meters would be six decades), we estimate that for
extended periods of time, meter-sized particles have surface mass density
$\sigma_L \approx 0.1 \sigma_{sol}$ where $\sigma_{sol} \sim 10^{-2} \sigma_g$
is the total surface mass density of solids at some location. We will define 
$\sigma_L/\sigma_{sol} \equiv f_L \approx 0.1$ as a key element of the model
described in section 2. 

\subsection{Radial drift: loss or transformation?}
The nebula gas, in general, has an outward pressure gradient, which counteracts
solar gravity to a small degree; the ratio of these two forces is $\eta \approx
H^2/R^2 \approx 2 c^2/\gamma V_K^2 \approx 2 \times 10^{-3} $, where $V_K$ is
the local Keplerian velocity (Weidenschilling 1977, Nakagawa et al 1986, Cuzzi
et al 1993) and $\gamma = 1.4$ is the ratio of specific heats. The gas orbits
more slowly than the solids at any location, and the ensuing headwind on
particles causes them to drift radially inwards at velocities which depend on
their size.  Even weak turbulence ensures that local particle
densities are too low to affect their drift velocities (Nakagawa et al 1986,
Cuzzi and Weidenschilling 2004). Typical radial drift velocities are shown in
{\bf figure 1}; they are strongly dependent on particle size, but only weakly
dependent on distance $R$ from the Sun for nebula models such as that adopted
($\sigma_g = 1700$(1 AU/$R$) g cm$^{-2}$). Meter-sized particles experience the
full headwind and are the most rapidly drifting; smaller particles experience a
smaller headwind, and larger particles  have increasing mass per unit area
(Weidenschilling 1977). For comparison we show the range of gas advection
velocity $V_n$ at 5 AU and $\alpha=10^{-3}$, which characterizes a range of
radial disk density profiles. The drift velocity $V_L \approx \eta V_K$ of
m-size particles is orders of magnitude larger than $V_n$. It is usually
inferred that such drifters are ``lost into the sun" on fairly short
timescales; however, this is not necessarily their fate --- as we describe
below.

\begin{figure}
\centerline{\psfig{figure=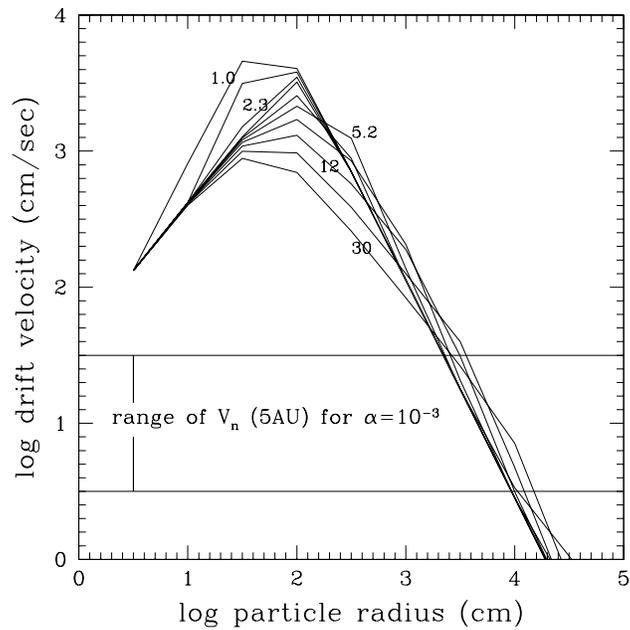,width=3.4 in,height=3.4 in}}
\caption{Radial drift velocity as a function of particle size, at a number of
nebula locations differing by a factor of 1.5 (several labeled, in AU from the
sun). The drift velocity at the peak is $\sim \eta V_K$. For comparison, a
range is shown for the nebula advection velocity $V_n$ at $R=$5AU with
$\alpha=10^{-3}$ (the range is associated with different nebula density
profiles; {\it cf. } Cuzzi {\it et al.} 2003).
\vspace{0.75 in}}
\end{figure}

When particles drift across the location where the midplane temperature exceeds
the sublimation temperature of one of their constituent species, that species
will evaporate locally within distance $\Delta R \sim (m/\dot{m})V_L$ where $m$
is the particle mass and $\dot{m}$ its evaporation rate. For water, $\Delta R <
$ 1 AU (Supulver and Lin 2000). Cyr et al (1998) have also modeled evaporation
of drifting water particles, finding that they drift considerable distances
before evaporating. However, these results are incompatible with those of
Supulver and Lin (2000) and with our own estimates using the same vapor
pressure expressions as Cyr et al. Even after discussions with J. Lunine
(personal communication, 2003) we cannot determine the cause of this
discrepancy. For silicates, $\Delta R$ is probably larger (Cuzzi et al 2003).
This {\it evaporation front} effect can produce significant enhancement of
material. We have considered only volatiles which represent significant
fractions of the total condensible mass at their evaporation
front: common iron-magnesium silicates, at about 1400K (Cuzzi et al 2003)
and water ice, at about 160K (this paper). The process is sketched in {\bf
Figure 2} and described further in section 2.

\begin{figure}
\centerline{\psfig{figure=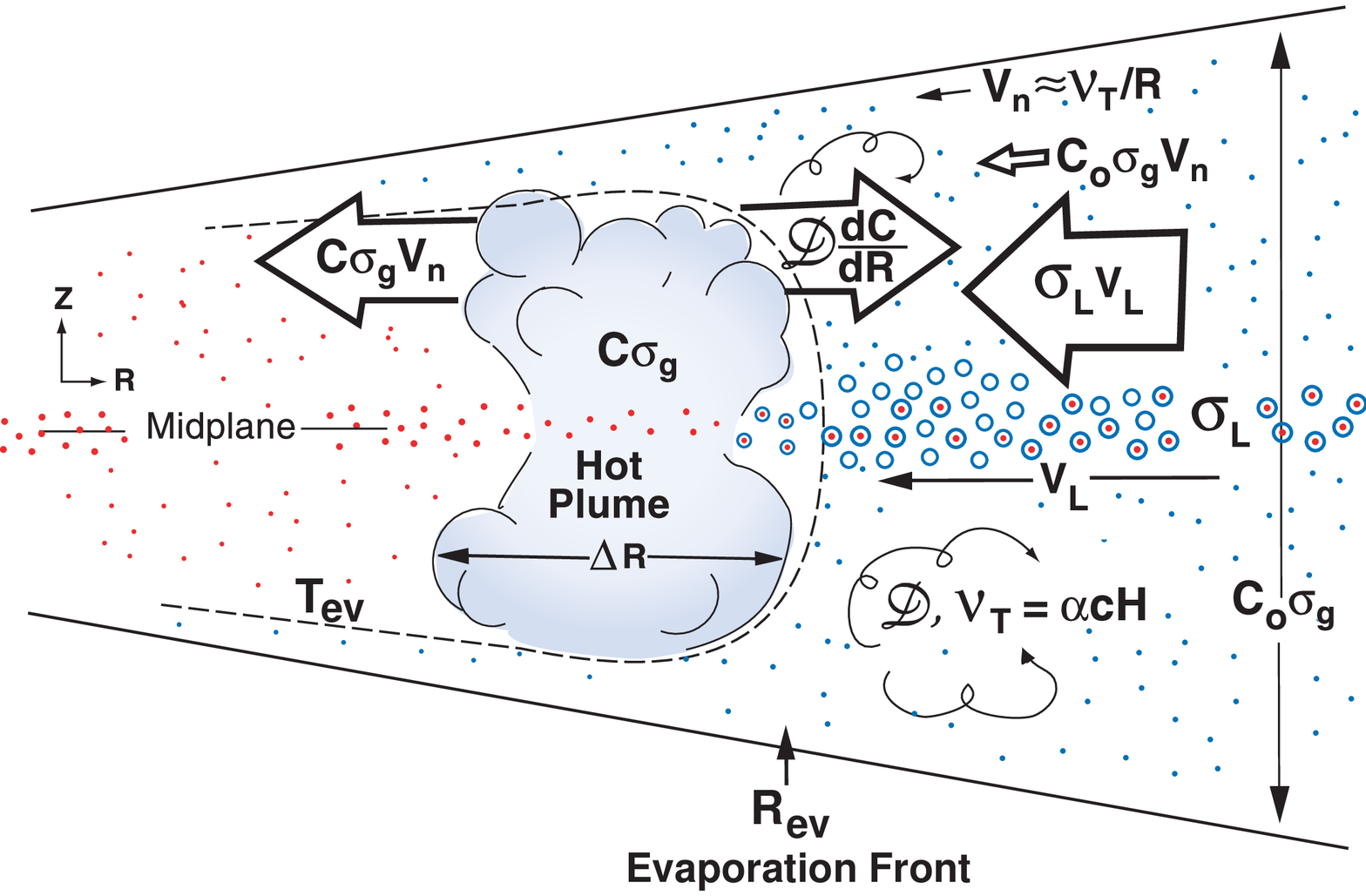,width=5.5in,height=3.5 in}}
\caption{Sketch illustrating inwardly drifting (blue) volatile material
crossing its evaporation front $R_{ev}$, with midplane temperature $T_{ev}$. The
surface density of large midplane solids is $\sigma_L= f_L C_o \sigma_g$; the
large inward drift flux of this material, $\sigma_L V_L$, can't be offset by
vapor removal processes $C \sigma_g V_n + {\cal D} \sigma_g dC/dR$ until the
concentration of the blue vapor $C$ is much greater than nominal solar $C_o$.
The more refractory (red) material, here shown as a minor constituent, simply
goes on drifting and growing.
\vspace{0.5 in}}
\end{figure}

\subsection{Caveats on assumptions}

The key parameter $f_L$ is uncertain. First, as discussed in section 1.2, and
most recently by Cuzzi and Weidenschilling (2004), we believe that the actual
combination of nebula turbulence $\alpha$ and particle strength (Sirono and
Greenberg 2000) allows particles to grow to meter-size. If this is not true, the
appropriate value of $V_L$ will decrease in proportion to the particle
size-density product (Cuzzi and Hogan 2003). This would decrease the mass flux
in ``large" particles. Secondly, one of our simplifying assumptions was that
growth was truncated above meter-size, and that planetesimals don't grow. This
led to our estimate of $f_L$ - the mass fraction in meter-sized particles - as
simply the inverse of the number of decades of size in the ``rubble" population
between microns and meters. If turbulence is vanishingly small and
planetesimals do grow, this simple logic loses its appeal. However, even in a
planetesimal growth regime, collisional erosion or breakup will continue to
occur. In this regime, $f_L$ might be regarded as the collisionally generated,
small-size end of a mass distribution with equal mass per decade, extending to
1000km.

\section{Model for evaporation fronts}
Here we will construct a simplified 1-D model that illustrates the basic
principles that govern transport of a condensible in the solar nebula. Let $C$
represent the mobile fraction of a condensible species in the nebula. We will
define this as the column density of the species (vapor or solid) divided by
the column density of the   nebula (chiefly hydrogen and helium gas). We
exclude from $C$ material that has condensed on large planetesimals;\ large
bodies will be treated as a stationary sink.

We further subdivide $C$ between the mass fraction $f_L$ in fast-drifting m-size
rubble (section 1.2) and the complementary fraction $1-f_L$ in small grains (or
vapor) that are strongly tied to the motions of the gas. In general we expect
that both collisional growth and disruption of particles are faster processes
than advection, so that the relative proportions of small grains to m-sized
rubble is roughly constant where solids are stable. The evaporation front is
defined by $R=R_{ev}$; inside the evaporation front $f_L=0$. For reasons
discussed in sections 1.2 and 1.4, we will assume that $f_L=0.1$ when $R\geq
R_{ev}$. Transport of $C$ is then described by
\begin{equation}\label{continuity}
{\partial \over \partial t}\Bigl\{\sigma_g C(R,t)\Bigr\} -
    {1\over R} {\partial \over \partial R} \Bigl\{(R\Phi(R,t)\Bigr\}
= -{f_L \sigma_g C\over \tau_{acc}}
\end{equation}                                                               
where the inward radial mass flux $\Phi$ is the sum of nebula advection,
diffusion, and particle drift, respectively:
\begin{equation}\label{flux} 
\Phi =  \left(1-f_L \right) C \sigma_g V_n
  + {\cal D}  {\partial \over \partial R}
   \Bigl\{ \left(1-f_L \right) C \sigma_g \Bigr\}
  + f_L C \sigma_g V_L .
\end{equation}
In equation \ref{continuity} the term on the right hand side represents the
accretion sink onto large planetesimals; $\tau_{acc}$ represents an accretion
time.  For reasons discussed above (see also Cuzzi and Weidenschilling 2004),
the m-size particles are much more concentrated toward the midplane than 
smaller size ranges which contain equal amounts of mass. We have therefore
assumed that accretion onto large bodies is dominated by m-size particles.

\subsection{Steady state solutions and a likely transient case}

We simplify equations \ref{continuity} and \ref{flux} by assuming
steady state, constant coefficients, and cartesian geometry.
The latter two assumptions introduce quantitative errors on the order
of unity provided that nebular properties (other than those associated
with the evaporation front) vary smoothly on the scale of $R$,
as is usually assumed in discussions of the nebula. We are left with
\begin{equation}\label{continuity2}
{d \over d R} \left\{ \left( 1 - f_L \right) C V_n
    + \left( 1 - f_L \right) {\cal D}
    {d C \over d R} + f_L C V_L \right\}
   = { f_L C \over \tau_{acc} }
\end{equation}
where as before $f_L=0$ for $R<R_{ev}$.

We then further simplify equation \ref{continuity2} by placing all accretion
onto planetesimals at the condensation front. This captures the spirit of the
snowline without introducing a complete model of planetary accretion. With this
simplification, equation \ref{continuity2} is directly solved analytically.
There are four boundary conditions. At large distances $R\gg R_{ev}$ the
nominal cosmic abundance is $C=C_o$; i.e. $C(R\rightarrow \infty) = C_o$. At
small distances $R\ll R_{ev}$ there is no source of $C$. This precludes the
purely mathematical solution in which outward diffusion balances inward
advection for $R<R_{ev}$. Consequently $C(R<R_{ev})=\overline{C}$, a constant.
The other two BCs apply at $R_{ev}$.  We assume that $C$ is continuous across
$R_{ev}$, and we apply a flux jump condition across $R_{ev}$,
\begin{equation}\label{flux_jump}
\Delta \Phi   = \Phi(R>R_{ev}) - \Phi(R<R_{ev})
= \int\limits^{R_{ev}+\delta R}_{R_{ev}-\delta R} {f_L C \sigma_g dR 
                                                    \over \tau_{acc}}
\equiv {\cal L} V_n \overline{C} \sigma_g,
\end{equation}
where ${\cal L}$ is a dimensionless sink factor integrated over the narrow band
of planetesimals just outside of $R_{ev}$, defined to make the sink term
similar in form to other terms in equation (3). The steady state solution that
results is
\[  C(R<R_{ev}) = E C_o  \]
\begin{equation}\label{solution}
  C(R>R_{ev}) = C_o \left[ 1 +
    \left(E-1\right) e^{\left\{-k\left(R-R_{ev}\right)\right\}} \right]
\end{equation}
where
\begin{equation}\label{beta}
   k \equiv { \left(1-f_L\right) V_n + f_L V_L \over \left(1-f_L\right) D }
\end{equation}
and
\begin{equation}\label{enhancement}
   E \equiv { \left(1-f_L\right) V_n + f_L V_L \over \left(1+{\cal 
L}\right) V_n }.
\end{equation}
The factor $E$ is the enhancement over cosmic abundance. If $f_L \ll 1$,
\begin{equation}\label{ezero}
E \approx { \left( 1 + f_L V_L /V_n \right) \over (1 + {\cal L}) }
= { E_o \over 1 + {\cal L}},
\end{equation}
where the factor $E_o$ is that of Cuzzi et al. (2003). 

In steady state the entire region interior to $R_{ev}$ is enhanced over cosmic
abundance (in the vapor) by the factor $E$. The distance scale $1/k$ is closely
related to $E$. It is like a skin depth. It represents the distance scale
beyond $R_{ev}$ in which solids are enhanced. Note that if $Pr_t<1$, the skin
depth deepens accordingly.

Enhancements can be large.  Using $V_L \sim \eta V_K$ we can estimate that, for 
$(f_L,{\cal L}) \ll 1$, 
\begin{equation}\label{Big_E}
  E \approx E_o \approx { f_L V_L \over V_n } \approx { 2 f_L \over 3 \alpha},
\end{equation}
which, for $f_L=0.1$ and $10^{-6} < \alpha < 10^{-3}$, is a very large factor
indeed. By contrast, Morfill and V{\"o}lk (1984) got much smaller vapor phase
enhancements (never exceeding solar) because they assumed particles which drift
only at about the same rate their nebula was advecting ($V_L = V_n$), and
because of their choice of outer boundary condition (their equation B7).

We discern three regimes of interest for $C(R,t)$, shown schematically in {\bf
figure 3}. Regimes 2 and 3 are the steady state solutions described by equation
\ref{solution} above. In regime 2 (${\cal L} \ll 1$), the entire region inwards
of $R_{ev}$ is enhanced by $E$ over solar. Regime 3 occurs when planetesimal
growth is significant and a sink appears at $R_{ev}$ (${\cal L} > 1$). If ${\cal
L}$ is big enough, the inner nebula can become depleted, essentially the result
of Stevenson and Lunine (1988). Regime 1 (sketched only conceptually in figure
3) is transient, because a certain amount of time is needed to reach steady
state. At first, evaporated material is found only within a radial band of
width $\Delta R$ (for water $\Delta R \ll R$, Supulver and Lin 2000). This
transient solution propagates toward the star and approaches a steady state
only after a time $t_{ss} \sim R_{ev}/V_n \approx 1/(3\pi \alpha \eta) \approx
40/\alpha$ orbit periods. Of course, $t_{ss}$ will also depend on $Pr_t$. For
$R_{ev}$=5 AU, $Pr_t = 1$, and $\alpha = 10^{-3}$ to $10^{-4}$, $t_{ss} \sim$ 0.5 -
5 Myr --- long enough to be interesting for the chemistry of the early inner
nebula (Cuzzi et al 2003). Depending on the rate at which ${\cal L}$ grows, the
nebula might evolve from regime 1 through regime 2 into regime 3, or directly
from regime 1 into regime 3.

\subsection{Global constraints on the model} 
Naturally, the steady state enhancement regime can't persist for the entire
duration of disk accretion. For example, in regime 2, with $E\sim 100$, as much
water is accreting onto the sun as hydrogen! This enhanced stage is limited in
duration and intensity by (a) growth of the planetesimal sink at $R_{ev}$,
leading to emergence of regime 3, and (b) depletion of the ultimate source of
the enhancement - outer solar system solids. 

In most nebula models, the surface mass density decreases as $1/R$; thus if the
nebula extended only to 50 AU, 10 times further than $R_{ev}$, and if {\it all}
the solids in that region were to be carried into the region interior to
$R_{ev}$, only an enhancement factor of 10 could be achieved; with $f_L <1$,
the limit could be even lower. However, the true radial extent and mass
distribution in the actual nebula are unknown; many protoplanetary disks are
not tens, but hundreds of AU across. Furthermore, some nebula models ({\it eg.}
Ruden and Pollack 1991) show the nebula surface mass density {\it increasing}
outwards, due to the effects of radially varying viscosity. In such a case, the
same global constraint allows enhancement by a factor of 400, even if the
nebula only extended to 50 AU. The likely time-variable nature of more
realistic solutions should be kept in mind. Even while global source
constraints limit the steady state solutions, large $E_o$ might prevail over
limited times and radial distances. Improving astronomical observations of the
radial extent and surface mass distribution of protoplanetary nebulae will be
helpful in establishing such global constraints. Overall, we do not feel that
values of $E_o \sim 10-100$ are unreasonable (especially in regime 1) but $E_o$
could be smaller (especially in regime 2) because of global constraints.

\begin{figure}                                                                 
\centerline{\psfig{figure=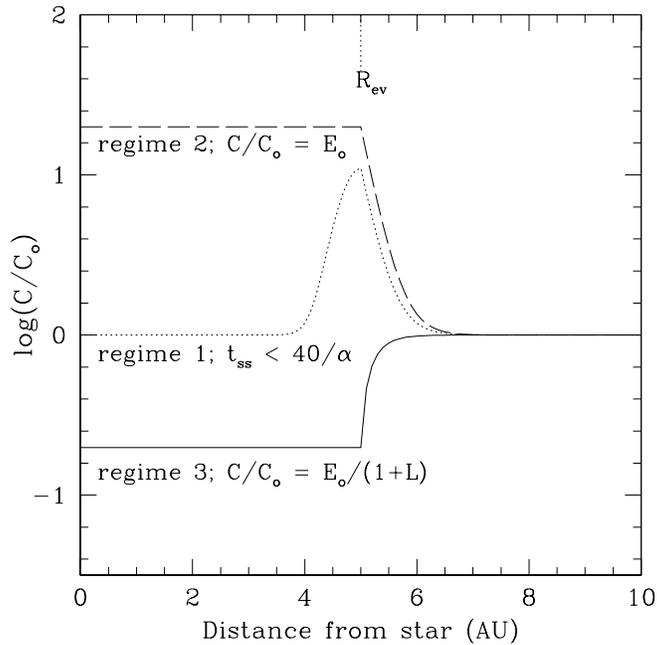,width=3.5in,height=3.5 in}}
\caption{Schematic of the radial (and temporal) variation of enhancement
$C/C_o$ for ``water" with an evaporation boundary at $R_{ev}$=5AU, taking for
illustration $E_o$=20. In regimes 1 and 2, there is no sink at $R_{ev}$
(${\cal L}=0$); regime 1 (dotted; schematic only) represents the transient
situation, where the inner nebula retains $C/C_o=1$ for typically $40/\alpha$
orbit periods. Regime 2 (dashed) is the steady state solution for ${\cal L}=0$.
As time proceeds and planetesimals grow in the enhanced solid density outside
$R_{ev}$, ${\cal L}$ increases; regime 3 (solid) illustrates the steady state
solution for $E_o$=20 and ${\cal L}=100$.  \vspace{0.5 in} }
\end{figure}

\subsection{The sink: planetesimal growth outside the condensation / evaporation
boundary} The sink term ${\cal L}$ removes solid material from further radial
evolution by accreting it onto the surfaces of immobile planetesimals just
outside $R_{ev}$. Thus the local mass density of potential planet-forming
objects increases. A detailed study of this process is well beyond the scope of
this paper, but the following simple expressions illustrate the possibilities.

The mass lost to the planetesimal sink can be written as
\begin{equation}\label{Mdot_PL_1}
\dot{M}_{PL} = 2 \pi R_{ev} E C_o \sigma_g V_n {\cal L}  = {\dot M} E C_o {\cal L}.
\end{equation}
Independently we
can write the mass accreted by a narrow belt of $N_{PL}$ planetesimals with
radius $r_{PL}$ and total mass $M_{PL}$ as
\begin{equation}  \label{Mdot_PL_2}        
\dot{M}_{PL} = N_{PL} \pi r_{PL}^2 \left( {f_L E C_o \sigma_g \over h_L}
\right) \Delta V \xi,
\end{equation}
where $\Delta V \sim V_L \sim \eta V_K$ is the relative velocity of sweepup,
$\xi$ is a sticking coefficient, and where we have ignored gravitational
focusing (appropriate for $r_{PL}<$ 30 km). 
Random velocities for meter-sized particles are comparable to $V_g$, so
$h_L/H \sim (\alpha/Pr_t)^{1/2}$ (Cuzzi and Hogan 2003; Cuzzi and Weidenschilling 
2004). Setting equations \ref{Mdot_PL_1} and \ref{Mdot_PL_2} equal to each
other, some algebra leads to
\begin{equation} \label{Cal_L}     
{\cal L} \approx \left({ M_{PL} \over 4 \pi R_{ev} H r_{PL} \rho_s} \right)
            {\xi f_L Pr_t^{1/2} \over \alpha^{3/2} }
             \approx 2 \times 10^{-6} \xi \alpha^{-3/2}
    		\left({ M_{PL} \over M_{\oplus}} \right)
                 \left({ {\rm 1 km} \over r_{PL}} \right) .
\end{equation}          
In evaluating equation \ref{Cal_L} we assume $f_L=0.1$, $\rho_s=1$, $Pr_t=1$, and
$R_{ev}=5$ AU.  Smaller bodies, which present a larger surface area for a given
mass, are more efficient sinks provided that they are large enough to be
immobile (greater than 100 meters or so; the size is itself
$\alpha$-dependent). Without detailed accretion modeling, it is hard to go
further. However, at this level of description, interesting ranges of values
for ${\cal L}$ and $\dot{M}_{PL}$ can be estimated {\bf (Table 1)}.

Equation \ref{Mdot_PL_1} may be rewritten to estimate the planetesimal belt
growth time (in orbit periods):
\begin{equation} \label{forty}       
{M_{PL} \over \dot{M}_{PL}}
= {M_{PL} \over 6\pi^2 R_{ev}^2 E C_o \sigma_g \eta \alpha {\cal L}}
  \approx { 2.4 \over E  \alpha {\cal L}}
\left({ M_{PL} \over M_{\oplus}} \right)
   \approx 40
\left({ M_{PL} \over M_{\oplus}} \right)
\left({ 1 + {\cal L} \over {\cal L}} \right).
\end{equation}
Equation \ref{forty} assumes a snowline at $R_{ev}=5$ AU, where $\sigma_g=300$
g/cm$^2$ and $C_o=0.01$. We used $E \sim E_o/(1+{\cal L})$ for $f_L \ll 1$, and 
equation (9) with $f_L = 0.1$.

In the ${\cal L} \ll 1$ regime, equation \ref{forty} may be combined with
equation \ref{Cal_L} to obtain a characteristic growth time for the belt (in
orbit periods, shown in line 3 of table 1):
\begin{equation} \label{twelve}   
{M_{PL} \over \dot{M}_{PL}} ({\cal L} \ll 1)
\approx 2 \times 10^7 \; \alpha^{3/2} \xi^{-1} \left({r_{PL} \over 
{\rm 1 km} }\right)
\end{equation}     
The transition from regime 2 to regime 3 can happen very quickly once nebula
turbulence dies down ($\alpha$ decreases) near $R_{ev}$; however, the poorly
understood sticking coefficient $\xi$ enters into all these timescales.

\begin{figure}
\centerline{\psfig{figure=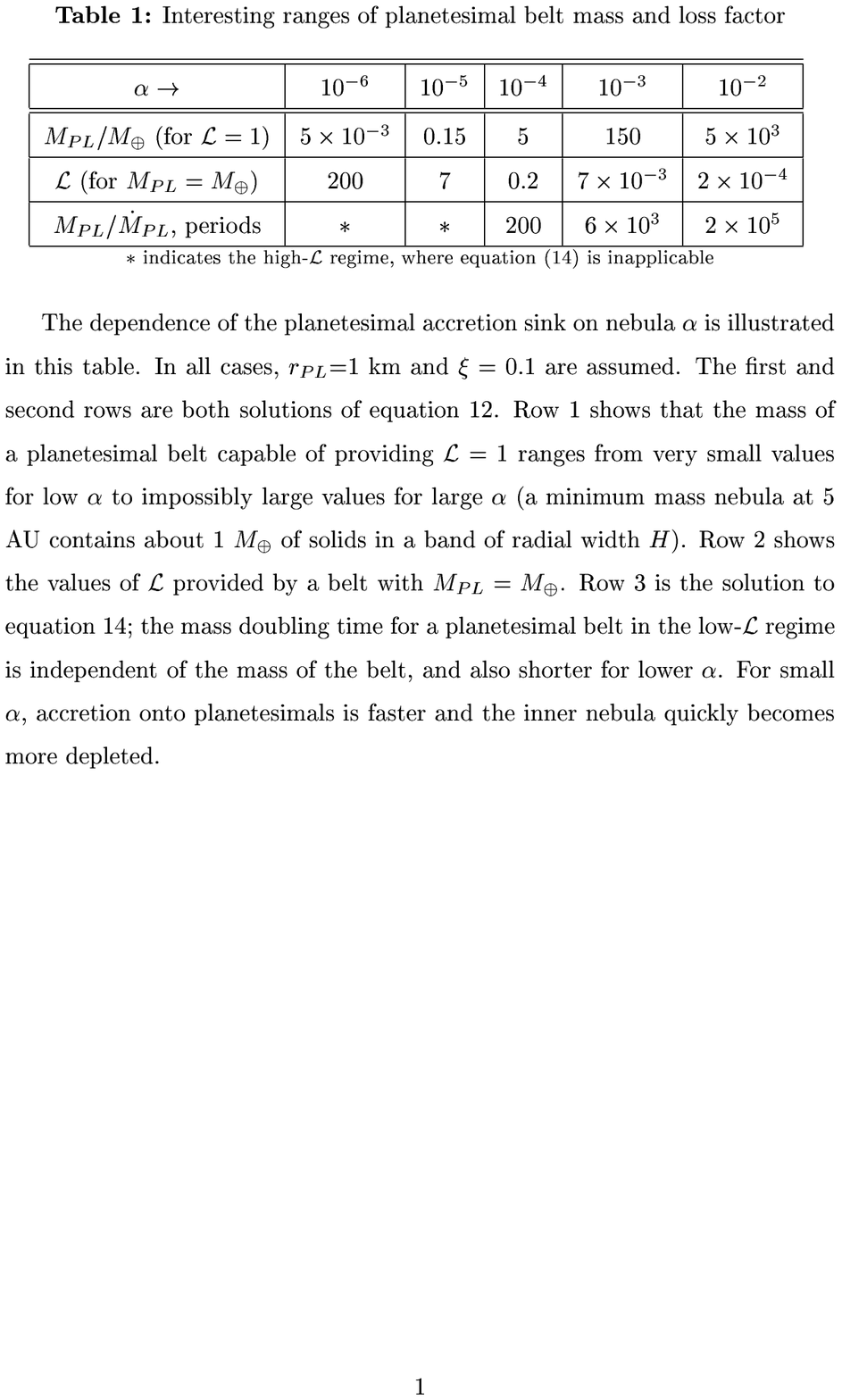,width=\textwidth,height=\textheight}}
\end{figure}

\section{Applications}
\subsection{Meteoritics and astronomical observations:}
Nebulae which are hot enough to evaporate silicates near the midplane in the
terrestrial planet region ($\dot{M}\sim10^{- 7} M_{\odot}$/yr, or age $\sim
10^5$ years; Bell et al 1997) are probably also young enough that the drifting
solids are more primitive and carbon-rich than chondrites. Evaporation of
silicates in the presence of 20-30\% carbon by number may lead to the formation
of abundant CO, with interesting mineralogical and isotopic implications (Cuzzi
et al 2003). For the duration of regime 1, this enhanced silicate and CO vapor
plume near $R_{ev}$(silicates) need {\it not} be accompanied by a similarly
enhanced component of water, because the water stripped out of the drifting
solids at $R_{ev}$(water) remains at radii $\gg R_{ev}$(silicates) until steady
state is achieved some $t_{ss}=40/\alpha$ orbit periods later (regime 2). The
potential duration of this dry, CO-and-silicate-rich inner solar system regime
1 seems to be comfortably longer than the apparent duration of the CAI
formation era in the inner solar system (Cuzzi et al 2003), which plausibly
ends when the inner nebula cools to below the evaporation temperature of common
silicates  - and thus long before the chondrule era that apparently occurs 1-3
Myr later ({\it eg.,} Amelin et al 2002; Russell et al. 2004).

Before planetesimal growth at $R_{ev}$(water) creates a sink, regimes 1 and 2
can provide an enhanced abundance of H$_2$O relative to hydrogen over a wide
range of locations. This may help explain several aspects of chondrite chemistry
indicative of elevated oxygen abundance, such as high levels of FeO in matrix
olivines in both ordinary and carbonaceous chondrites (Nagahara 1984, Scott et
al 1984, 1989; Wood 1988). It has traditionally been argued that enhancement of
nebula gas in silicates of chondritic composition can provide the high oxygen
fugacity required for high-FeO silicates to form in the nebula (Palme and
Fegley 1990). Recently, however, Fedkin and Grossman (2004) have shown that
chondritic silicates are ineffective in this regard because they provide too
much S (which competes for Fe) in addition to their O. Our mechanism 
enriches the nebula gas in H$_2$O alone, and might provide the needed
oxygen fugacity without the sulfur complications. In another application,
Ciesla et al (2003) have suggested that fine grained silicates can be aqueously
altered in the nebula {\it gas} by shock waves, if the nebula gas is enhanced
in H$_2$O by something like a factor of 100. This level of enhancement is
achievable, even if perhaps only regionally or for a limited time, under
circumstances described here. Furthermore, the enhancement is probably
temporally and spatially variable, depending on how the nebula evolves between
the three regimes we have identified. In addition, A. Krot (2003, private
communication) has noted that this enhancement can affect the Oxygen isotopic
ratios in primitive meteorite minerals in a time-variable way, if outer solar
system ice has different O-isotopic composition from inner solar system
silicates (eg., Lyons and Young 2004; Yurimoto et al 2004; Krot et al. 2004).

Some recent astronomical observations seem to show abundant CO in the
terrestrial planet regions of vigorously accreting protoplanetary nebulae
(Najita et al 2003). The presence of abundant CO might be associated with the
evaporation front of primitive silicate-carbon material discussed above. In at
least one case, the water content in the inner nebula seems to be low relative
to CO (Carr et al 2003). This could be the signature of regime 1, or perhaps a
very early stage regime 3. Future observations of this type, perhaps at higher
spatial resolution,  might help us determine evolutionary timescales and
connect the current properties of external protoplanetary nebulae with the
record of the accretion process in our own.

\subsection{Planetary formation} Formation of Jupiter has long been
associated with the concept of a snowline (Wuchterl et al 2000). The nominal
scenario for the formation of an icy jovian core in less than a few Myr, while
the nebula gas is still present, requires that the surface mass density of
solids in the formation region exceed that of a minimum mass nebula by almost
an order of magnitude (Lissauer 1987). One well-known proposal for this
enhancement is the cold finger effect, in which the entire water content of the
{\it inner} nebula is diffusively transported to the snowline and frozen out
there (Stevenson and Lunine 1988). With the assumptions of a vigorously
turbulent inner nebula, a narrow condensation annulus, and no leakage back into
the inner solar system (questioned by Sears 1993), the cold finger effect
leads to an enhancement of solids outside the snowline by a factor
of 6-25 in about $10^5$ years. This can be expressed as a mass
flux of roughly $\pi R_{ev}^2 C_o \sigma_g / (R_{ev}^2/{\cal D}) \sim \pi C_o
\sigma_g {\cal D} \sim$ a few times $10^{-5}$ Earth masses per year for $Pr_t
\sim 1$. The ratio of the mass flux to $R_{ev}$ due to solids
drifting from {\it outer} regions (this paper), to that of solar abundance
vapor diffusing from the {\it inner} solar system is
\begin{equation}
{ 2 \pi R_{ev} f_{L} C_o \sigma_g V_L \over \pi C_o \sigma_g {\cal D}}
= {2 R_{ev} f_L V_L \over {\cal D}} \approx 3 E_o Pr_t 
\approx { 2 f_L \over (\alpha/Pr_t)}.
\end{equation}
Unless the turbulent Prandtl number is very small, {\it i.e}.\ unless turbulent
transport is much larger than viscous transport, and given the validity of our
particle size distribution arguments, inward particle drift and vapor retention
would seem to be the dominant source for enhancement of solids near $R_{ev}$.

Looking somewhat further beyond the boundaries of this paper, we suspect that
evaporation fronts of low-temperature volatiles might also have important
implications for their enhancement in the gaseous envelope of Jupiter, a
problem highlighted by Owen et al (1999) and Atreya et al (2003). Another
possibility, associated with $R_{ev}$(silicates), is the isotopic
homogenization of a large amount of the silicate material which ultimately ends
up in meteorite parent bodies. The gross isotopic homogeneity of meteoritic
silicates has been a persistent puzzle, because few nebula models evaporate
silicates throughout the asteroid formation region. It has also been suggested
to us (J. Chambers, personal communication, 2003) that the process, operating at
$R_{ev}$(silicates), might help explain the mass distribution in the
terrestrial planet formation region. Some of these applications will be
addressed in future papers.

{\it Summary:} We show that nebula constituents will be enhanced in the vicinity
of their condensation/evaporation boundaries $R_{ev}$, due to rapid inward
drift of solid material in the form of meter-sized boulders, and slow
subsequent removal of the ensuing vapor. This {\it evaporation front} effect
modifies not only the surface mass density of solids available just {\it
outside} $R_{ev}$ (useful for planet building), but also the chemistry and
mineralogy of material which resides well {\it inside} $R_{ev}$ (of potential
importance to meteoritics and gas giant atmospheres). The enhancements can be
one or even two orders of magnitude, and probably vary on timescales of a
million years or so - perhaps also exhibiting significant radial variation
during that time. Some of these properties might be observable by astronomical
observations.

\section*{Acknowledgements:} We thank Stu Weidenschilling and Robbins Bell for 
helpful conversations during the course of this research. We thank Lynne 
Hillenbrand for drawing our attention to the astrophysical observations, and 
Joan Najita and John Carr for helpful discussions and preprints in advance of 
publication. We thank Larry Grossman for discussions regarding the 
applicability to meteoritic silicates, Sasha Krot for discussions regarding 
Oxygen isotope applications, and John Chambers for the suggestion that 
there may be an application to mass densities in the terrestrial planet region. 
We thank Jack Lissauer and John Chambers for reviews of an early draft. We
thank Rich Young for bringing the Jovian volatiles problem to our attention.
We thank our reviewer, David Stevenson, for useful suggestions on presenting
caveats. This research was supported by a grant to JNC from NASA's origins of
Solar Systems Program and a grant to KJZ from NASA's Exobiology Program.

\section*{References}
\begin{list}{}{\leftmargin \parindent \itemindent -\parindent \itemsep 0in}

\item Amelin, Y., A. N. Krot, I. D. Hutcheon, and A. A. Ulyanov (2002); Science
297, 1678

\item Atreya, S. K.; Mahaffy, P. R.; Niemann, H. B.; Wong, M. H.; Owen, T. C.
(2003); Planet. Sp. Sci. 51,  105

\item Bell, K. R., Cassen, P. M., Klahr, H. H., and Henning., Th. (1997); 
Astrophys. J. 486, 372

\item Calvet, N., Hartman, L. and Strom, S. E. (2000); in Protostars and
Planets, V. Mannings, A. P. Boss, and S. S. Russell, eds; Univ. of Arizona
Press, 377

\item Carr, J. S., A. T. Tokunaga, and Najita, J. (2004) Astrophys. J. 603, 213

\item Ciesla, F., Lauretta, D. S., Cohen, B. A., and Hood, L. L. (2003)
Science, 299, 549

\item Cuzzi, J. N., S. S. Davis, and A. R. Dobrovolskis (2003) Icarus, 166, 385

\item Cuzzi, J. N., A. R. Dobrovolskis, and J. M. Champney (1993)
Icarus, 106, 102

\item Cuzzi, J. N., R. C. Hogan, J. M. Paque, and A. R. Dobrovolskis (2001)     
Astrophys. J., 546, 496

\item Cuzzi, J. N., and S. J. Weidenschilling (2004); in Meteorites and the
Early Solar System II; D. Lauretta, L. A. Leshin, and H. McSween, eds; Univ. of
Arizona Press (submitted).

\item Cyr, K. Sears, W. D., and Lunine, J. I. (1998) Icarus, Volume 135, Issue
2, pp. 537-548

\item Dubrulle, B., G. E. Morfill, and M. Sterzik (1995); Icarus 114, 237

\item Fedkin, A. V. and L. Grossman (2004); 35th LPSC, paper \#1823.

\item Krot, A. N. et al (2004) in preparation

\item Lissauer, J. J. (1987); Icarus 69, 249

\item Lyons, J.R. and Young, E.D.  (2004), 35th Lunar and Planetary Science
Conference (LPI, Houston) abstract no. 1970

\item Morfill, G. E. and H. J. V{\"o}lk (1984) Astrophys. J. 287, 371

\item Nagahara, H. (1984) Geochim. Cosmochim. Acta 48, 2581

\item Najita, J., J. S. Carr, and R. D. Mathieu (2003); 
Astrophys. J. 589, 931

\item Nakagawa, Y., M. Sekiya, and C. Hayashi (1986); Icarus 67, 375

\item Owen, P.R. Mahaffy, H.B. Niemann, S.K. Atreya, T. Donahue, A. Bar-Nun   
and I. de Pater (1999) Nature 402 (1999), 269

\item Palme, H. and Fegley, B. (1990) Earth Planet. Sci. Lett. 101, 180

\item Prinn, R. G. (1990); Astrophys. J. 348, 725

\item Ruden, S. and J. B. Pollack (1991) Astrophys. J. 375, 740

\item Russell, S. S., L. Hartmann, J. Cuzzi, A. N. Krot, M. Gounelle, and S.
Weidenschilling (2004); in ``Meteorites and the early solar system -- II", D.
Lauretta, L. A. Leshin, and H. McSween, eds; Univ. of Arizona Press (submitted).

\item Scott, E. R. D., A. E. Rubin, G. J. Taylor, and K. Keil (1984); 
Geochim. Cosmochim. Acta 48, 1741

\item Scott, E. R. D., D. J. Barber, C. M. Alexander, R. Hutchison, and J. A.
Peck (1989) in ``Meteorites and the early solar system", J. F. Kerridge and M.
S. Matthews, eds; Univ. of Arizona Press; p 718

\item Sears, W. D. (1993); 24th LPSC, p. 1271-1271.

\item Sirono, S. and J. M. Greenberg (2000) Icarus 145, 230-238

\item Stepinski (1998) Icarus, 132, 100

\item Stepinski, T. F. and P. Valageas (1997); Astronomy and Astrophysics, 319, 1007

\item Stevenson, D. J. and J. I. Lunine (1988); Icarus, 75, 146

\item Stone, J. M., C. F. Gammie, S. A. Balbus, and J. F. Hawley (2000); 
in Protostars and Planets IV; p589-599; V. Mannings, A. P. Boss, and S. S.
Russell, eds. Univ. of Arizona Press

\item Supulver,  K. amd Lin, D. N. C. (2000); Icarus 146, 525

\item Weidenschilling, S. J. (1977) Mon. Not. Roy. Ast. Soc. 180, 57

\item Weidenschilling, S. J. (1989); in ``Meteorites and the early solar
system", J. F. Kerridge and M. S. Matthews, eds; Univ. of Arizona Press; p 348

\item Weidenschilling, S. J. (1997) Icarus 127, 290

\item Weidenschilling, S. J. (2000); Sp. Sci. Rev. 92, 295

\item Weidenschilling, S. and J. N. Cuzzi (1993); in ``Protostars and Planets
III"; E. Levy and J. Lunine, eds; University of Arizona Press

\item Wood, J. A. (1988); Ann. Revs. Earth Planet. Sci., 16, 53

\item Woolum, D., and P. M. Cassen (1999) Meteoritics Planet. Sci. 34, 897

\item Wuchterl, G., T. Guillot, and J. J. Lissauer (2000); in Protostars and
Planets IV; eds V. Mannings, A. P. Boss, A.P., and S. S. Russell, University of
Arizona Press; p. 1081

\item Yurimoto, H. and K. Kuramoto (2004) Science, in press

\end{list}

\end{document}